\documentclass[preprint,showpacs,floats,nofootinbib,color,showkeys]{revtex4}
\usepackage[dvips]{graphicx}

\newcommand{\bea}{\begin{eqnarray}}
\newcommand{\eea}{\end{eqnarray}}
\newcommand{\be}{\begin{equation}}
\newcommand{\ee}{\end{equation}}
\newcommand{\ba}{\begin{eqnarray}}
\newcommand{\ea}{\end{eqnarray}}

\begin{document}

\title{Superscaling analysis of the Coulomb Sum Rule in quasielastic 
electron-nucleus scattering}

\author{
J.A. Caballero$^1$, 
M.C. Mart\'{\i}nez$^2$, 
J.L. Herra\'{\i}z$^2$, 
J.M. Ud\'{\i}as$^2$
}
\affiliation{$^1$Departamento de F\'{\i}sica At\'omica, Molecular y Nuclear,
Universidad de Sevilla, 41080 Sevilla, SPAIN \\
$^2$Grupo de F\'{\i}sica Nuclear, Departamento de 
F\'{\i}sica At\'omica, Molecular y Nuclear,
Universidad Complutense de Madrid, 28040 Madrid, SPAIN}

\date{\today}


\begin{abstract}

The Coulomb sum rule for inclusive quasielastic electron
scattering in $^{12}$C, $^{40}$Ca and $^{56}$Fe is analyzed based on
scaling and superscaling properties. Results obtained in the
relativistic impulse approximation with various descriptions of the
final state interactions are shown. A comparison with experimental
data measured at Bates and Saclay is provided. The theoretical description
based on strong scalar and vector terms present in the relativistic mean 
field, which has been shown to reproduce the experimental asymmetric 
superscaling function, leads to results that are in fair agreement with
Bates data while it sizeably overestimates Saclay data. We find that the 
Coulomb sum rule for a momentum transfer $q\geq 500$ $MeV/c$ saturates to a value
close to 0.9, being very similar for the three nuclear systems considered. This 
is in accordance with Bates data, which indicates that these show
no significative quenching in the longitudinal response.

\end{abstract}


\maketitle

\section{Introduction}

Strongly correlated many-body systems are of interest in very diverse areas of physics. In particular,
nuclei have been explored in depth by means of electron scattering reactions for very different
kinematical situations. More than 50 years of experimentation have proved that electron
scattering provides one of the best tools for investigating the structure of nuclear systems
and their constituents~\cite{Fru84,Bof96,Kel96,Ras89,Walecka01}.
The electromagnetic interaction with which electrons probe 
nuclei is well under control and is weak enough so that the process can be treated in first order
photon exchange. Under this assumption (Born Approximation), it is possible to isolate the different 
components of the nuclear response by changing appropriately the electron kinematical variables.
Indeed, assuming the Plane Wave Born Approximation (PWBA), i.e., only one virtual photon exchanged and 
electrons described as free particles, the inclusive $(e,e')$ quasielastic (QE) differential 
cross section is written as~\cite{Fru84,Bof96,Kel96},
\be
\frac{d\sigma}{d\varepsilon'd\Omega'}
=\sigma_M\left[v_LR^L(q,\omega)+v_TR^T(q,\omega)
\right]\, ,
\label{eq1}
\ee
where $(\varepsilon',\Omega')$ are the energy and solid angle of the
scattered electron, $\sigma_M$ is the Mott cross section, and $v_L(v_T)$ the
longitudinal (transverse) leptonic kinematic factors that in the extreme relativistic limit 
(ERL) for the electrons are simply given as $v_L=(Q^2/q^2)^2$ and $v_T=\tan^2(\theta_e/2)-Q^2/2q^2$
with $\theta_e$ the electron scattering angle and $(\omega,q)$ the energy and momentum
transferred in the process ($Q^2=\omega^2-q^2$).
The hadronic $R^K(q,\omega)$ response functions are constructed from the
nuclear electromagnetic tensor $W^{\mu\nu}$ given in terms of
the initial and final many-body nuclear state, and the
nuclear electromagnetic many-body current operator~\cite{Bof96,Kel96,Walecka01,Ras89}.

In the case of QE kinematics and the momentum of the exchanged photon large enough
(its wavelength being of the order of or smaller than the nucleon size),
knockout of a single nucleon is the dominant contribution to the nuclear response.
Under these conditions, the impulse approximation (IA) holds, and the 
inclusive $(e,e')$ cross section can be given as the integrated semi-inclusive 
single-nucleon knockout cross sections. This approximation, which is implicit in scaling
analyses, has been shown to work properly in the kinematic region dominated by
the QE process~\cite{Chinn89,Jourdan,Udias,highp,Udias3}. 

Lowest nucleon resonances are mainly excited with purely transverse photons, hence they
do not affect the longitudinal response which essentially captures the purely 
nucleonic contribution to the nuclear response. 
Assuming that the nucleons are the only relevant degrees of freedom, sum rules have been
derived in both relativistic~\cite{Chinn89} and nonrelativistic 
schemes~\cite{McVoy65}. These
sum rules can be stated separately for the longitudinal and the
transverse contributions to the inclusive cross section. In particular, 
the Coulomb sum rule (CSR) states that by integrating the longitudinal strength over the full range of
energy loss $\omega$ at large enough momentum transfer $q$, one should get the total charge 
(number of protons) of the nucleus. While the experimental realization of the transverse sum rule 
gets contributions from resonance excitations and thus, will likely be above theoretical
estimates based only upon nucleonic degrees of freedom, the experimental CSR
is suitable to comparison with theoretical predictions. Not only the asymptotic value of the 
CSR for large $q$, but also the evolution of the CSR with increasing $q$, is of interest 
in order to test nuclear models and/or descriptions of the reaction mechanism.

Indeed, enormous experimental efforts have been made at different laboratories, 
Saclay~\cite{Saclay1,Saclay2,Morg01},
Bates~\cite{Bates}, JLAB~\cite{CSR-JLAB}, to get separated longitudinal and transverse contributions from
QE electron scattering data.
The analysis of data and its impact on the CSR for different nuclei have been discussed in the literature, 
leading to different conclusions concerning the role played by several ingredients:
nucleon correlations, final state interactions, modification of the nucleon form factors 
by the nuclear medium, etc. Jourdan concluded that the integrated longitudinal 
($L$) response function saturates 
for $q$ high enough at the 100\% of the CSR limit~\cite{Jourdan}, 
and thus it is not suppressed, showing no $A$-dependent quenching. On the contrary, 
from the analysis of data taken at Saclay~\cite{Saclay1}, Morgenstern and Meziani
have concluded the existence of a significant quenching of the CSR, 
and have interpreted such suppression as due to the change of the nucleon 
properties inside the nuclear medium~\cite{Morg01}.
Being aware of the present controversy,
the most comprenhensive effort to measure separated longitudinal inclusive 
responses from several nuclei and different values of momentum transfer $q$, 
in a large enough range of energy and with unprecedent high statistics and small systematic errors, 
has been recently completed at JLAB~\cite{CSR-JLAB}. This experiment is 
currently under analysis and preliminary results will be released in the next months.

An alternative procedure to get some insight into the CSR relies on the 
information provided by the scaling properties of the longitudinal separated data. As already
shown in previous works~\cite{Day90,DS199,MDS02}, world $(e,e')$ data have clearly demonstrated 
the validity of scaling and superscaling (independence of the scaled response on the kinematics
and on the nuclear target) behavior. In particular, the analysis of the
separated $L$ contribution has led to introduce a {\sl ``universal''}
superscaling function, which contains the relevant information about
the initial and final state nuclear dynamics explored by the probe~\cite{MDS02}. 
Superscaling was originally introduced within 
the simple Relativistic Fermi Gas (RFG) model that, albeit showing
perfect scaling and superscaling properties, yields
a superscaled function shape not in accordance with data~\cite{MDS02,scaling88}.

The experimental superscaling function has an asymmetric shape, with a 
long tail exhibiting strength for energy transfers well beyond the RFG domain.
Further, most nonrelativistic models also lack the significative 
strength at high-$\omega$~\cite{neutrino2}.
The presence of this tail is of relevance for the CSR analysis, 
as the sum rule requires integration of the strength in the whole energy transfer range 
(up to infinity), which is of course not feasible from the experimental point of view.
The integration of the experimental strength ends at some finite value 
of the transferred energy,
located where the asymmetric tail of the superscaling function resides.
Thus, what is left out of the integration region from theoretical estimates of the CSR 
would highly depend on whether the model does or does not reproduce this asymmetric tail.
The main aim of this work is trying to shed some light to the CSR problem, 
making use of the experience acquired during the analysis of the scaling and 
superscaling phenomenon~\cite{DS199,MDS02}.

\section{Scaling and Superscaling} 

The usual procedure to get the scaling function consists in dividing the
inclusive differential cross section (\ref{eq1}) by the
appropriate single-nucleon $eN$ elastic cross section, weighted by the 
corresponding proton and neutron numbers~\cite{DS199,MDS02,Barbaro:1998gu} 
involved in the process,
\be
f(\psi',q)\equiv k_F\frac{\left[\displaystyle\frac{d\sigma}{d\varepsilon'd\Omega'}\right]}
{\sigma_M\left[V_LG_L(q,\omega)+V_TG_T(q,\omega)\right]} \, .
\label{fscaling}
\ee
$\psi'(q,\omega)$ is the dimensionless scaling variable extracted
from the RFG analysis that incorporates the typical momentum scale for the selected
nucleus~\cite{MDS02}.
The fully relativistic expressions for $G_{L(T)}$ involve proton and neutron 
form factors $G_{E(M)pn}$, weighted by proton and neutron numbers, and
an additional dependence on the nuclear scale given through the Fermi momentum $k_F$
(explicit expressions are given by Eqs.~(16-19) in ~\cite{MDS02}).
Analogously, the analysis of the separated longitudinal ($L$) and transverse ($T$)
contributions leads to scaling functions,
\be
f_{L(T)}(\psi',q)\equiv k_F\frac{R^{L(T)}(q,\omega)}{G_{L(T)}(q,\omega)} 
\label{fltscaling} \, .
\ee

At transferred energies above the QE peak, scaling is violated 
in the transverse channel by effects beyond the impulse approximation~\cite{Day90,DS199}.
However, the available data for the $L$ response are compatible with scaling in all 
the QE region. This has made it possible to extract 
an experimental scaling function $f_L^{exp}$, that effectively represents the 
nucleon contribution to the nuclear response under QE kinematics~\cite{DS199,MDS02}.
In this work we use the Relativistic Impulse Approximation (RIA) that 
leads to a hadronic tensor evaluated from the transition 
single-nucleon current matrix elements. These are constructed making use of the relativistic bound-state, 
the scattering wave function and the relativistic single-nucleon electromagnetic current operator.
We guide our analysis with calculations where the bound nucleon states correspond to self-consistent
Dirac-Hartree solutions, derived within a relativistic mean field (RMF) 
approach. The outgoing nucleon wave function is given as a solution of the
Dirac equation in presence of a relativistic potential which takes care of
the final state interactions (FSI) between the ejected nucleon and the residual
nucleus. In previous works~\cite{PRL05,jac06,Chiara03} it has been investigated the
role played by different descriptions of FSI: i) the use of the same relativistic
mean field employed to describe the initial bound states and ii) considering the
phenomenological relativistic optical potential derived by Clark {\it et al.}~\cite{Clark}
but with their imaginary part set to zero in order to consider all final 
channels and not only the elastic one,
that we denoted as rROP. Finally, ignoring all distortion from FSI leads to the relativistic plane 
wave impulse approximation (RPWIA) where the knocked out nucleon is treated as a plane wave.

\begin{figure}[htb]
\begin{center}
\includegraphics[scale=0.8]{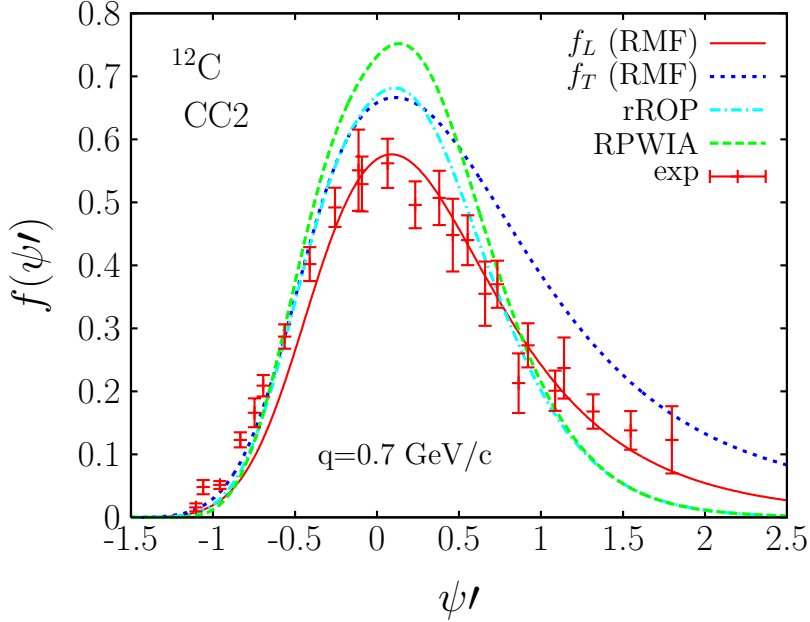}
\caption{(Color online) Superscaling function for $^{12}C(e,e')$ evaluated with the RPWIA, 
rROP and RMF approaches compared to the experimental function. In the RMF case,
separate $L$ and $T$ contributions are shown. 
}
\label{fLT_scaling}
\end{center}
\end{figure}

In recent works~\cite{PRL05,jac06,Amaro07} it has been shown that the RMF model, 
where the same relativistic potentials
are applied to the initial and the final state, reproduces satisfactorily the magnitude and detailed 
shape of $f_L^{exp}$, while other models fail to reproduce the long tail appearing at high energy 
transfer $\omega$ (large positive values of the scaling variable $\psi'$). 
This is clearly illustrated in Fig.~\ref{fLT_scaling} where we present 
the superscaling function evaluated within
the RIA and different descriptions of FSI. Results correspond to $^{12}$C and $q=700$ MeV/c.
The scaling function obtained using the real part of
the relativistic optical potential (rROP) is compared to the plane wave limit (RPWIA) and
to results obtained in presence of the scalar and vector terms in the relativistic mean
field potential (RMF), where the separate $L$ and $T$ contributions to the scaling function are
plotted. Scaling of zeroth kind, {\it i.e.,} $f_L=f_T=f$, 
has been shown to be fulfilled within RPWIA and rROP approaches~\cite{jac06,jac_plb}. 
In all the cases, the CC2 prescription for the current operator has been 
selected~\cite{jac06}. As observed, $f_T$ obtained within RMF is increased 
with regard to $f_L$. This is due to the off-shellness of the nucleons, 
modest for RPWIA and rROP and much more important for RMF because of the
stronger potentials involved in the final nucleon states. While the function $f_L$ 
hardly changes (a consequence of current conservation), $f_T$
exhibits a significant dependence with off-shell nucleon effects~\cite{jac06,jac_plb}.

It is worth mentioning that correlations also shift strength towards larger energy values, 
as they allow for multi-nucleon knockout. Correlations have been a common ingredient
of theoretical predictions of CSR~\cite{omar1,omar2,omar3} and can also explain the asymmetrical
shape of the superscaling function~\cite{BCS}. In this work our focus is not the
explanation of the observed asymmetry, that has been discussed in previous 
work~\cite{PRL05,jac06,Amaro07,jac_plb}, but rather explore its effect on the
predicted CSR values. The comparison with the experimental $L$ superscaling function,
also provided in Fig.~\ref{fLT_scaling}, shows that the RMF approach 
follows closely the behavior of data describing also the asymmetrical shape 
of $f_L^{exp}$.

Moreover, the RMF model, as studied in previous work~\cite{jac06, jac_plb}, 
fullfills the continuity equation and dispersion relations, hence being adequate to inclusive 
scattering where all nucleon propagation channels, not only the elastic one described by 
the optical potentials, must be incorporated. The different behavior presented by 
RMF and rROP (Fig.~\ref{fLT_scaling}) is linked directly to the strong potentials present 
in the RMF for large values of the energy transfer. On the contrary, rROP potentials tend to 
weaken significantly with increasing energy values. This is also consistent with 
the similar behavior shown by rROP and RPWIA results. The potentials modify the effective 
values of the momenta at the nucleon vertex, giving rise to a shift of strength to 
(asymptotical) larger values of $\omega$ (see Refs.~\cite{PRL05,jac06,Amaro07,jac_plb}).
Finally, use of relativistic optical potentials (with imaginary term) in inclusive
reactions has been applied within the relativistic Green's function (GF) approach, leading
to similar results to RMF, {\it i.e.,} with presence of the asymmetry in the scaling
function for intermediate $q$-values~\cite{pavia}.

\section{The Coulomb sum rule}

Including relativistic corrections~\cite{Forest} and the structure
of the nucleons, the explicit expression for the CSR, widely used by
experimentalists in the analysis of the separate $L$-data~\cite{Saclay1,Bates}, 
is written as
\be
CSR(q)=\frac{1}{Z}\int_{\omega^+}^\infty \frac{R^L(q,\omega)}{\widetilde{G}_E^2(Q^2)}d\omega 
\,  \label{CSR_1}
\ee
with the effective electric form factor given by
\be
\widetilde{G}_E^2(Q^2)=\left[G^2_{Ep}(Q^2)+\frac{N}{Z}G_{En}^2(Q^2)\right]\frac{(1+\tau)}{(1+2\tau)}
\, , \label{effective}
\ee
where $N$ and $Z$ are the neutron and proton numbers of the target, respectively, and
$G_{Ep}$ and $G_{En}$ the Sachs electric form factors for proton and neutron. The
term $\tau$ is the usual dimensionless quantity, $\tau\equiv |Q^2|/4M_N^2$
with $M_N$ the nucleon mass. The lower
limit in the integration $\omega^+$ includes all inelastic contributions but excludes the
elastic peak.


\begin{figure}[tb]
\begin{center}
\includegraphics[scale=0.63]{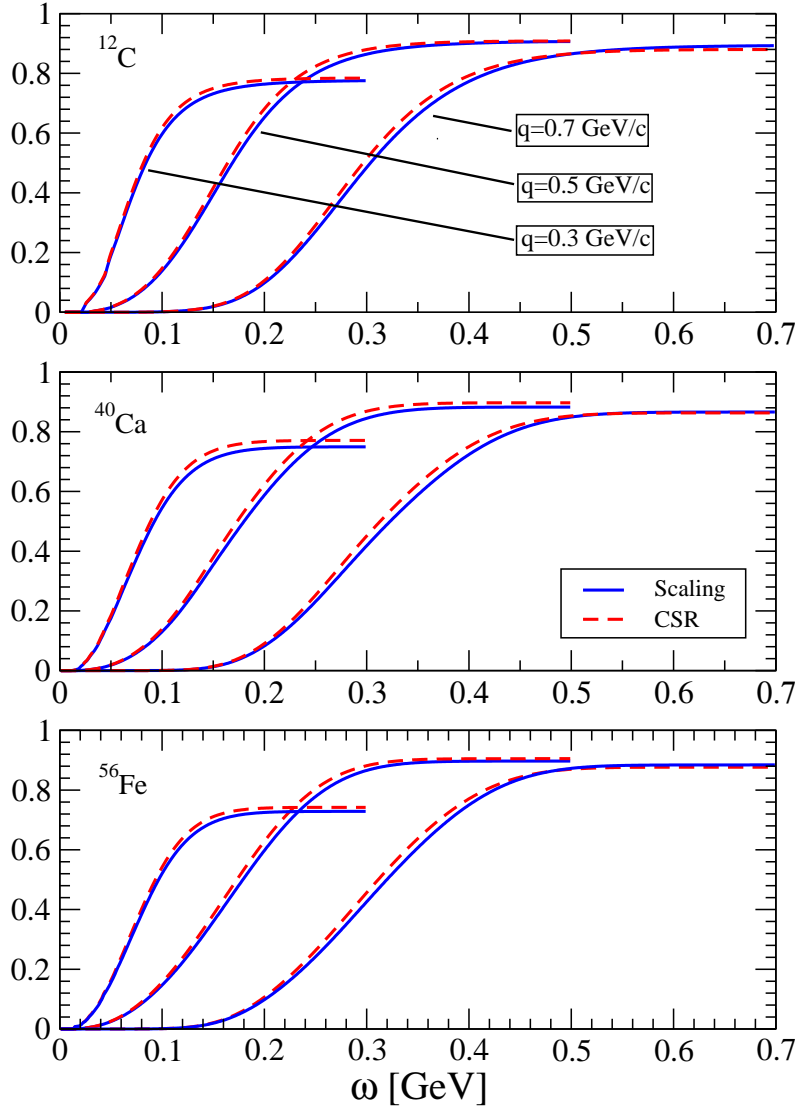}
\vspace{0.25cm}
\caption{(Color online) Coulomb sum rule as a function of the energy transfer for $^{12}$C (top panel),
$^{40}$Ca (middle) and $^{56}$Fe (bottom). In each case, results obtained using the expression of 
the CSR given by Eq.~(\ref{CSR_1}) are compared with predictions based on the scaling 
analysis (\ref{CSR_2}) for three different values of the momentum transfer.
}
\label{CSR1_vs_CSR2}
\end{center}
\end{figure}

%
%

An analog of the CSR can be also introduced in terms of the superscaled function 
and the scaling variable by taking
into account the explicit expression of the longitudinal superscaling function, 
as well as its physics significance,
\be
CSR_{scal}(q)=\int_{-\infty}^{\infty}d\psi' f_L(\psi') \, . \label{CSR_2}
\ee
Here the integration limits, denoted by $(-\infty,+\infty)$, extend in reality to the range 
allowed by kinematics and the experimental setup. Note that the scaling variable depends on the 
transferred momentum and energy, $q,\omega$. Expression (\ref{CSR_1}) of the CSR used by
experimentalists does not exactly correspond 
to Eq.~(\ref{CSR_2}) due to the fully relativistic expressions involved in
the longitudinal scaling function~\cite{MDS02} and to the different integration variable. 
Thus, in order to set down the impact of the particular CSR expression on the analysis of data, 
in what follows we compare results corresponding to Eqs.~(\ref{CSR_1}) and (\ref{CSR_2}). The analysis is
presented in Fig.~\ref{CSR1_vs_CSR2} where we have considered three nuclei: $^{12}$C (top panel),
$^{40}$Ca (middle) and $^{56}$Fe (bottom). In each case we show how the CSR behaves
as a function of the energy transfer $\omega$ for three different values of the momentum transfer
$q$: $0.3$, $0.5$ and $0.7$ GeV/c. We compare the results corresponding to Eq.~(\ref{CSR_1}),
denoted as CSR (dashed line), with the ones evaluated through the scaling function (\ref{CSR_2}),
denoted as Scaling (solid line). We conclude that, apart from some minor discrepancies ascribed
to the different single-nucleon expressions considered and the influence of the nuclear scale
introduced in the longitudinal scaling function, both expressions for the CSR lead to
similar results, hence drawing analogous conclusions. 
Notice that in all the cases the result given by Eq.~(\ref{CSR_2}) lies slightly 
below the one of (\ref{CSR_1}) for intermediate values of the energy transfer.
All results in Fig.~\ref{CSR1_vs_CSR2} have been obtained with the RIA-RMF model.

Comparing the results obtained for the three nuclei, the CSR dependence with the target is 
seen to be very tiny. The CSR saturates to almost the same value
for the three nuclei: $\sim 0.9$ for $q=0.5$ and $0.7$ GeV/c and $\sim 0.7$ for $q=0.3$ GeV/c.
Moreover, the behavior of the CSR is similar for the
three targets, getting saturation, at each $q$-value, for very close transferred energies.
Results in Fig.~\ref{CSR1_vs_CSR2} allow us to focus on the 
CSR predictions given by Eq. (\ref{CSR_2}) and to compare them to data arranged according 
to~(\ref{CSR_1}).


\begin{figure}[tb]
\begin{center}
\includegraphics[scale=0.54,angle=270]{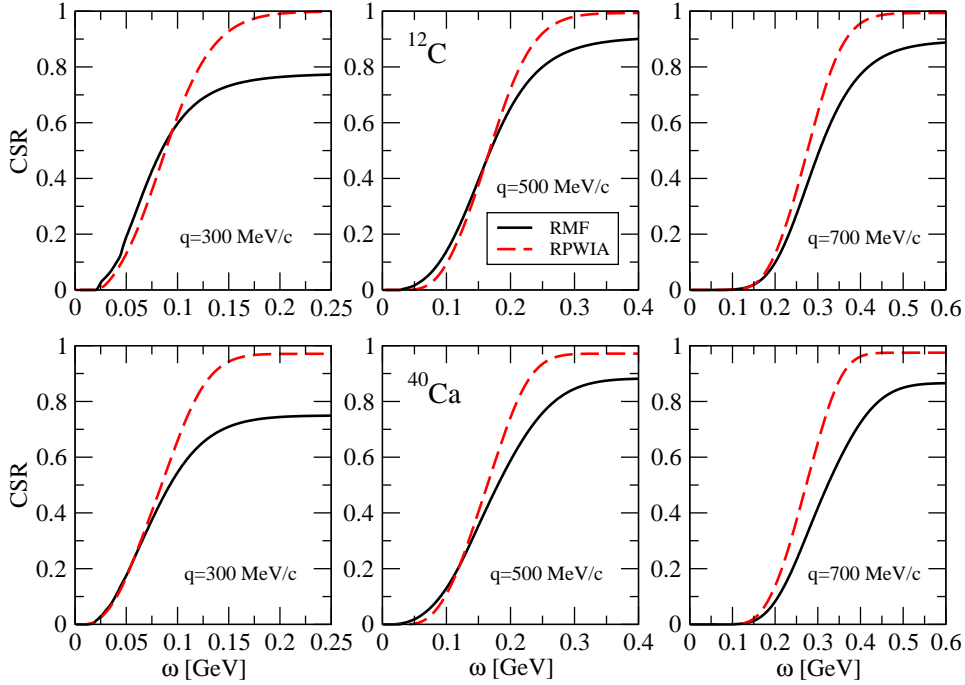}
\vspace{0.65cm}
\caption{(Color online) Coulomb sum rule according to Eq.~(\ref{CSR_2}) for $^{12}$C (top panels) 
and $^{40}$Ca (bottom) vs the energy transfer $\omega$. Results
are shown for three values of the momentum transfer $q$, comparing RPWIA (dashed) and RMF
(solid) approaches. 
}
\label{CSR_model}
\end{center}
\end{figure}

%
%

As shown in Fig.~\ref{fLT_scaling}, the function $f_L^{exp}(\psi')$ presents
a long tail extended to large $\omega$-values, which is not reproduced by
RPWIA and rROP calculations (neither by the majority of
nonrelativistic models employed in the literature~\cite{neutrino2}).
The presence of this important strength in $f_L^{exp}(\psi')$ may affect significantly
the results for the CSR. Hence, in what follows we study how the CSR depends
on the specific approach considered. To make easier the analysis we only consider two extreme
cases: RPWIA, namely no FSI, and RMF, {\it i.e.,} the presence of strong scalar and vector
potentials in the final state. The shape of $f_L(\psi')$ in both cases is quite different,
being the tail at large $\omega$-values largely absent in RPWIA.

Results are presented in Fig.~\ref{CSR_model} for $^{12}$C (top panels) and $^{40}$Ca (bottom) and
three $q$ values: $q=300$ MeV/c (left panels), $500$ MeV/c (middle) and 700 MeV/c (right). 
One observes that the CSR saturates to $\sim 1$ for all $q$-values and 
the two nuclei in the case of RPWIA. 
On the contrary, the RMF description leads to a saturation value smaller than 1, which 
grows with $q$ up to being of the order of $0.9$ for $q\geq 0.5$ GeV/c, {\it i.e.,} 
where Pauli blocking is 
not in effect and thus also scaling holds. It is also important to point out that saturation is reached 
faster in RPWIA. This is consistent with the general symmetry shown by $f_L^{RPWIA}(\psi')$ 
in contrast to the long tail 
presented by data and the RMF model (see Fig.~\ref{fLT_scaling}). 
Part of the strength that has been shifted to 
high $\omega$-values in the RMF case (because of FSI) cannot be reached within the kinematical constrains,
hence leading to RMF-CSR results being smaller than the RPWIA ones. 
Comparing the results obtained for $^{12}$C and 
$^{40}$Ca as well as $^{56}$Fe (not shown in the figure but following a very similar trend),
the CSR dependence on the target is very tiny (see discussion in previous figure). 
This outcome is in accordance with second kind scaling property.


\begin{figure}[htb]
\begin{center}
\includegraphics[scale=0.5,angle=270]{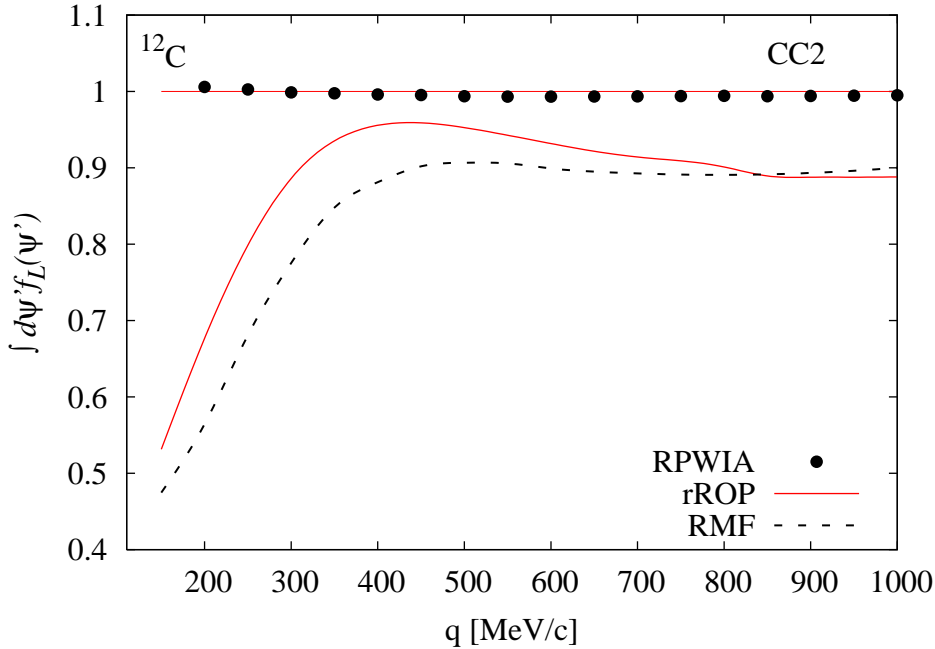}
\caption{(Color online) Integrated scaling function versus the momentum transfer $q$. Results are shown
for $^{12}$C comparing three different models, integrated up to the maximum
energy transfer allowed by kinematics (see text for details).
}
\label{CSR_q}
\end{center}
\end{figure}

The behavior of the CSR with the momentum transfer $q$ is illustrated in Fig.~\ref{CSR_q}, where
we show the results for $^{12}$C evaluated through Eq.~(\ref{CSR_2}) with different 
descriptions of the FSI: RPWIA, rROP and RMF, integrated up to the maximum energy transfer
allowed by kinematics. The CC2 current operator has
been considered, as results with this prescription
agree fairly well with the superscaling function. As observed, RPWIA leads 
to unity for all $q$-values, even in the region where the CSR is not 
saturated in the RFG due to Pauli blocking, 
{\it i.e.,} $q\leq 400-450$ MeV/c. 
This is in contrast with the other two models with FSI turned on, 
which show a CSR that increases with $q$ up to becoming stable. 
Concerning specific CSR-saturated values, the rROP gets its maximum ($\sim 0.95$) 
for $q=0.4$ GeV/c, starting to diminish slightly in the region $0.4\leq q\leq 0.9$ GeV/c up to being of the
order of $\sim 0.9$. The CSR result obtained within RMF increases with $q$ 
up to reaching $\sim 0.9$ for $q\approx 400-500$ MeV/c. This CSR value remains stabilized in the
whole $q$ region explored in the figure, that is, up to $q=1$ GeV/c.
For higher $q$, not presented in Fig.~\ref{CSR_q}, it can be shown that the CSR-RMF
(likewise rROP) result starts to increase approaching 1 for $q\sim 1.6$ GeV/c. However,
for so large $q$-values, caution should be drawn on the assumptions implied by our 
theoretical description as well as by the extraction of the CSR from data. Finally,
note that the strong potentials involved in the RMF, both for initial and final nucleon states, 
make the strength to be shifted to higher values of the (asymptotic) nucleon 
momentum~\cite{Amaro07}.


Comparison between CSR theoretical results and experimental data requires
to extract the Coulomb sum rule from the longitudinal response data 
by performing the integrals in Eqs.~(\ref{CSR_1}), (\ref{CSR_2}) using as upper
integration limits the specific $\omega$-cutoff values employed by the experimentalists.
In particular, in the case of $^{40}$Ca, different
experiments, Bates~\cite{Bates} and Saclay~\cite{Saclay1}, 
have considered different $\omega_{max}$-values 
as integration limits, as shown in Tables~\ref{table1} and~\ref{table2}. 
One has to keep in mind that in the case of Bates data and for $q\geq 425$ MeV/c,
the value of the maximum energy transfer included in the experimental CSR
diminishes as the momentum transfer $q$ goes up due to the 
uncertainties associated with the L/T separation. This explains why in Table~\ref{table1}, 
while the total CSR estimated under the RMF reaches a rather constant value ($\sim 0.88$) 
for transferred momenta larger than 425 MeV/c, the
predicted CSR under RMF employing the experimental energy transfer cutoff 
gets smaller for $q$ increasing, after $\sim 400$ MeV/c.  

\begin{table}[h]
\vspace{1cm}
\begin{center}
\begin{tabular}{cccccc} \hline \hline
\ \ $q$ [MeV/c] \ \   & \ \  $\omega_{max}$ [MeV] \ \  &\ \   CSR (total)\ \   & \ \  CSR($\omega_{max}$)\ \   & \ \  $\%$(diff.)\ \   &\ \  CSR(RFG) 
 \\ \hline
300 &  140  & 0.7493  &  0.6917  & 7.7 &  0.8197 \\
325 &  160  & 0.7889  &  0.7378  & 6.5 &  0.8640 \\
350 &  190  & 0.8207  &  0.7874  & 3.9 &  0.8975 \\
375 &  220  & 0.8458  &  0.8234  & 2.6 &  0.9289 \\
400 &  250  & 0.8638  &  0.8485  & 1.8 &  0.9483 \\
425 &  260  & 0.8759  &  0.8548  & 2.4 &  0.9649 \\
450 &  240  & 0.8822  &  0.8159  & 7.5 &  0.9683 \\
475 &  230  & 0.8842  &  0.7552  & 14.6 &  0.9707 \\
\hline
\hline
\end{tabular}
\end{center}
\caption{Integrated CSR evaluated within the RMF as a function of the momentum transfer $q$. Second
column presents the maximum energy loss as indicated in Bates experiment~\cite{Bates}. We
present the CSR results evaluated by extending the integration up to the maximum energy transfer allowed by 
kinematics (column 3) and up to the cutoff value used in Bates (column 4). Finally, column 5
reflects the difference between both results (percentage) and column 6 presents for
reference the RFG predictions.}
\label{table1}
\end{table}
%
\begin{table}[h]
\vspace{1cm}
\begin{center}
\begin{tabular}{ccccc} \hline \hline
\ \ $q$ [MeV/c]\ \   &\ \   $\omega_{max}$ [MeV]\ \   & \ \  CSR (total) \ \  & \ \  CSR($\omega_{max}$)\ \   & \ \  $\%$(diff.) 
 \\ \hline
330 &  175  & 0.7889  &  0.7586  & 3.8 \\
370 &  195  & 0.8458  &  0.7953  & 6.0 \\
410 &  235  & 0.8638  &  0.8387  & 2.9 \\
450 &  265  & 0.8822  &  0.8490  & 3.8 \\
500 &  290  & 0.8825  &  0.8335  & 5.6 \\
550 &  310  & 0.8788  &  0.8003  & 8.9 \\
\hline
\hline
\end{tabular}
\end{center}
\caption{Same as Table~\ref{table1}, but for the kinematics considered at Saclay~\cite{Saclay1}.}
\label{table2}
\end{table}

\begin{figure}[tb]
\begin{center}
\includegraphics[scale=0.5]{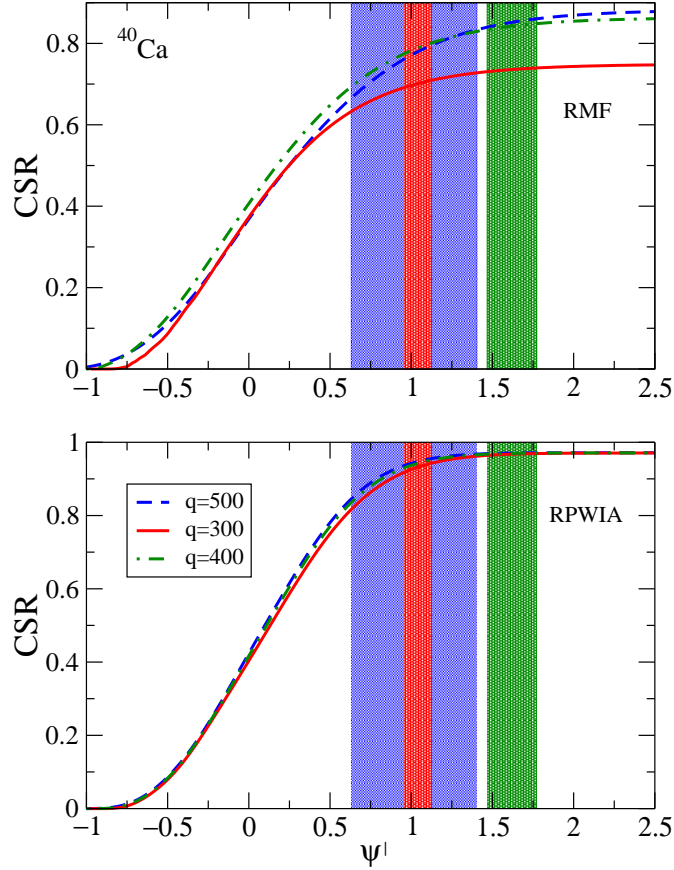}
\vspace{0.2cm}
\caption{(Color online) Coulomb sum rule as a function of the scaling variable $\psi'$ for $^{40}$Ca.
Top panel refers to results obtained within the RMF approach and bottom to RPWIA. 
The vertical shadowed bands refer to the extreme values
of $\psi'$ corresponding to the energy transfer cutoffs considered in the analysis of the experiments.
Each color refers to a different $q$-value, namely red ($q=300$ MeV/c), green ($q=400$ MeV/c) and
blue ($q=500$ MeV/c). Lower limits in each band correspond to Bates values and higher ones to Saclay
integration cutoff.}
\label{CSR_exp}
\end{center}
\end{figure}

In Fig.~\ref{CSR_exp} we present results for $^{40}$Ca corresponding to RMF
(top panel) and RPWIA (bottom) approaches. In each case, three $q$-values have
been considered, $q=300$, 400 and 500 MeV/c. The CSR is shown as a function
of the scaling variable $\psi'(q,\omega)$. We also plot, for each $q$, 
the value of the scaling variable $\psi'$ corresponding to the specific 
$\omega$-cutoffs given in the experimental papers. 
These span the regions: $140\lesssim\omega\lesssim 150$ MeV/c for $q=300$ MeV/c,
$230\lesssim\omega\lesssim 250$ MeV/c for $q=400$ MeV/c, and 
$220\lesssim\omega\lesssim 290$ MeV/c for $q=500$ MeV/c. In the latter ($q=500$ MeV/c) the
lower $\omega$-value represents the limit employed at Bates~\cite{Bates} 
and the larger one the cutoff included in Saclay~\cite{Saclay1}. 
These regions are presented as shadowed areas, where the color indicates the specific $q$-value
which is directly connected with the corresponding (same color) theoretical CSR result. 

Results in Fig.~\ref{CSR_exp} illustrate clearly the amount of saturation reached by the CSR 
at the maximum $\omega$-loss taken from the experiment. Let us consider the case $q=300$ MeV/c
(solid red line and red shadowed band). Here the CSR saturates to $\sim 0.75$ for RMF and 
$\sim 0.97$ for RPWIA if the integration is extended over the whole allowed range 
(see Table~\ref{table1}). 
On the contrary, CSR results integrated up to the shadowed area are approximately $\sim 0.70$ (RMF)
and $\sim 0.93$ (RWPIA). 
This means that saturation of CSR is reached at the order of $\sim 93\%$ 
for RMF and $\sim 96\%$ for RPWIA at  the experimental energy cutoff. In other words, 
the $\omega$-values beyond the
experimental accessible region correspond to a $\sim 7\%$ contribution 
to the fully integrated CSR in the RMF, and only $\sim 4\%$ in
the RPWIA case. These results reflect the increased tail of the longitudinal response in the RMF case.
It is worth recalling, however, the different CSR values emerging from the
the two approaches, $0.75$ in RMF and almost 1 ($0.97$) in RPWIA.

Similar comments apply also to higher $q$-values, $400$ MeV/c (green color) and $500$ MeV/c (blue),
although here the discrepancy between RMF and RPWIA results gets reduced because of the significant
enhancement of the CSR value in the RMF approach. For $q=400$ MeV/c, the RMF-CSR experimental 
cutoff result is on average $\sim 97\%$ of the RMF-CSR for the whole range, whereas in RPWIA saturation 
is already reached at the experimental energy loss. Finally, in the case of $q=500$ MeV/c some comments apply
because of the wide blue shadowed area linked to the very different $\omega$-cutoffs
considered at Bates and Saclay, $\omega_{max}=220$ MeV and 290 MeV, respectively. 
For the Saclay experiment~\cite{Saclay1}, {\it i.e.,} upper limit in the shadowed band, 
the CSR model evaluated up to the experimental cutoff includes 
$\sim 95\%$ ($\sim 100\%$) contribution of the total CSR strength in the RMF (RPWIA)
approach. On the contrary, the contribution (integrated up to $\omega_{max}$) reduces to
$\sim 75\%$ ($\sim 95\%$) for RMF (RPWIA) in the case of the maximum energy loss used at
Bates~\cite{Bates} (lower limit of the band, $\omega_{max}=220$ MeV). 
As it will be shown later on, this makes a significant difference when
comparing theoretical calculations with the CSR extracted from both experiments.

It is important to point out again that the CSR obtained in the whole allowed $\omega$ range 
within the RMF approach saturates to 
$\sim 0.88$ for $q\geq 400, 500$ MeV/c, that is, $\sim 12\%$
below the RPWIA result. Further the RMF-CSR result accumulated 
up to the experimental energy cuttof employed in Bates, 
is on average $\sim 15-18\%$ below the corresponding 
RPWIA result for $q=400,\ 500$ MeV/c, and $\sim 25\%$ below for $q$ around 300 MeV/c. 
This is consistent with the behavior shown in Fig.~\ref{CSR_q}.
It is worth noticing that the contribution to the CSR of the 
strength outside the experimentally integrated region is different if considering RPWIA and/or RMF. 
In a model like RMF which agrees with the experimentally deduced 
longitudinal scaling response of Bates, the contribution
of the unobserved tail beyond the cutoff employed at Bates is around 7\% for $q\sim 300$ MeV/c, 
2-3\% around $q\sim 400$ MeV/c and increases up to 15\% for the largest $q$-value 
(475 MeV/c) measured at Bates.

\begin{figure}[htb]
\begin{center}
\includegraphics[scale=0.6,angle=270]{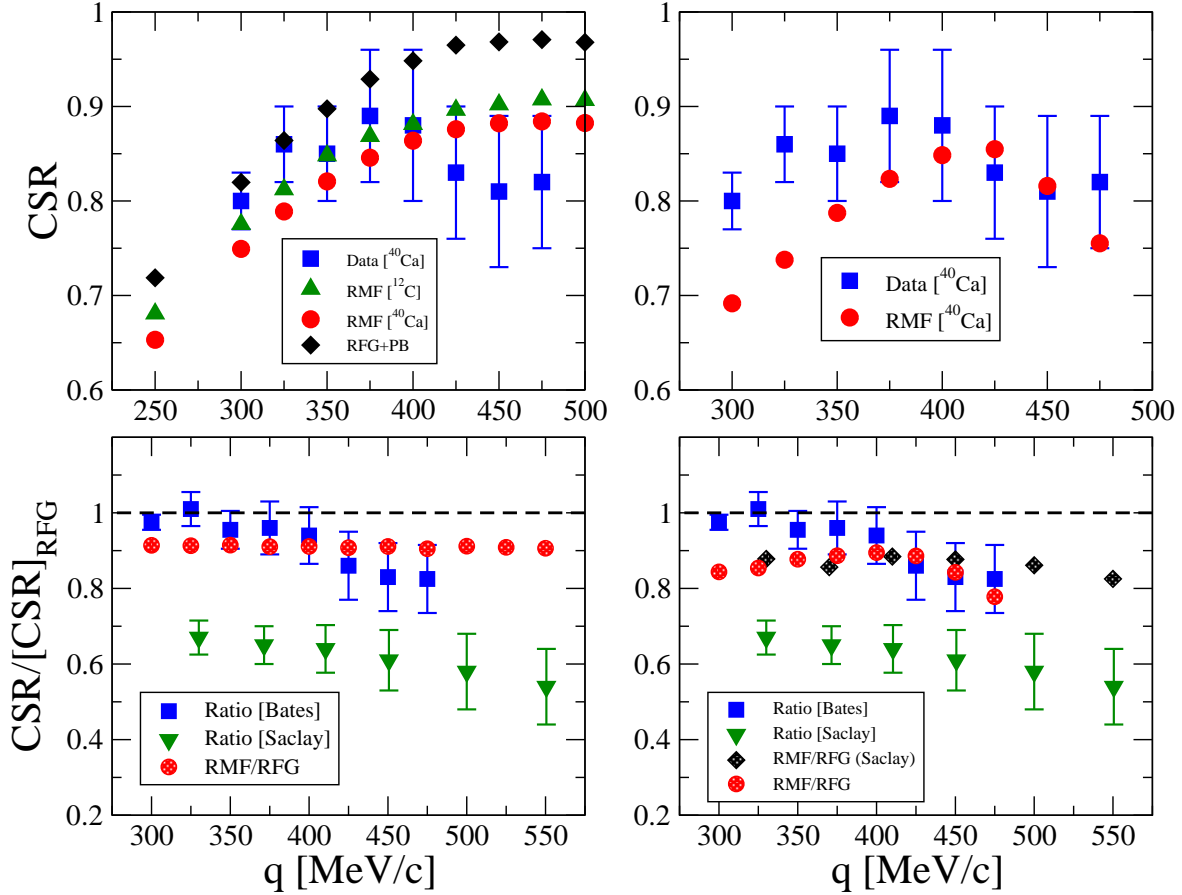}
\caption{(Color online) Coulomb sum rule compared to data. Top-left panel shows results obtained
with the RMF for $^{12}$C (green triangles) and $^{40}$Ca (red circles) and RFG with Pauli blocking 
(black diamonds). In all cases, integration in Eq.~(\ref{CSR_2}) has been extended to the whole
region allowed by kinematics. Theoretical results are compared with
data from Bates corresponding to $^{40}$Ca. Top-right panel: as in previous case but RMF calculations
evaluated using the $\omega$-cutoff values given in \cite{Bates}. 
Bottom panels present the ratio between RMF results
and RFG ones compared to data from Bates and Saclay (see text for details).
}
\label{theory_data}
\end{center}
\end{figure}

To conclude, a comparison between theory and experimental data is provided in
Fig.~\ref{theory_data}. First, in the left-top panel theoretical results for the CSR
evaluated with the RMF approach applied to $^{40}$Ca (red circles) and $^{12}$C
(green triangles) are presented. In both cases, CSR has been 
obtained making use of (\ref{CSR_2}) and extending the upper 
integration limit to the maximum value permitted by the kinematics, i.e.,
once CSR has already reached saturation. Results in Fig.~\ref{theory_data} 
show the independence of the CSR on the nuclear target, within the present approaches.
For reference, we also include 
the CSR evaluated with the RFG model (black diamonds). Here, 
the CSR result approaches almost 1 for $q\sim 500$ MeV/c, {\it i.e.,} $q\geq 2k_F$. For lower
$q$-values Pauli-blocking effects are important giving rise to a significant reduction in 
the CSR value. Notice that, although the integral of $f_L^{RFG}$ (likewise the CSR) 
should be exactly one in the QE domain, the value in Fig.~\ref{theory_data}, 
slightly lower than 1, reflects the shift energy included in the definition 
of the scaling variable $\psi'$ (see \cite{MDS02,jac06} for details).
Theoretical results are compared with the Coulomb sum rule for $^{40}$Ca
extracted from data measured at Bates~\cite{Bates} for $q$-values in the
domain, $300\leq q\leq 475$ MeV/c. On general grounds, we observe that
RMF results agree fairly well with data, lying slightly below for the
smaller $q$-values, $[300,350]$ MeV/c, and above data for $q>400$ MeV/c 
where the experimental uncertainty is significantly larger. 
Notice however, that the behavior shown by data, with
a depletion occurring for $q\geq 400$ MeV/c is not reproduced by theoretical
CSR calculations, which increase smoothly with $q$ approaching saturation.
This discrepancy is mainly linked to the upper integration $\omega$-limits used in the
analysis of data. Whereas theoretical CSR results were obtained through
Eq.~(\ref{CSR_2}) extending the integral up to the maximum $\omega$,
likewise $\psi'$, value permitted by kinematics, Bates CSR data on the contrary,
have been extracted
making use of Eq.~(\ref{CSR_1}) with the upper $\omega_{max}$ limit fixed, for
each $q$, according to the values given in Table II of 
Ref.~\cite{Bates} (see also Table~\ref{table1}). In particular, notice the relatively
low $\omega_{max}$ values used by experimentalists for $q \sim 450,500$ MeV/c. 
As we have already mentioned, significant strength in the CSR may be left out when using 
relatively low energy transfer cutoffs. 

This is clearly illustrated in the right-top panel of Fig.~\ref{theory_data}, where
we compare again Bates CSR data with RMF theoretical results, but these now evaluated using
as upper integration limits the same $\omega_{max}$ values considered in the experiment
(see Table~\ref{table1} in this work and Table II in~\cite{Bates}). Compared with previous results, 
a decrease in the RMF-CSR is observed,
depending its magnitude on the specific momentum transfer considered: from $\sim 2-4\%$ for the
central $q$ ($[350-425]$ MeV/c) and $\sim 6-8\%$ for $q=300, 325, 450$ MeV/c,
up to $\sim 15\%$ for $q=475$ MeV/c.
This explains the depletion presented by the CSR 
(theory and data) for larger $q$. Concerning the comparison between theory and data,
we observe that Bates CSR data are reproduced within the RMF approach. Only for
$q=300$ and $325$ MeV/c, RMF results underestimate data by $\sim 10-12\%$. This 
discrepancy can be partly ascribed to the different expressions used to evaluate
the CSR, Eq.~(\ref{CSR_1}) for experimental data and (\ref{CSR_2}) for RMF. 
As shown in Fig.~\ref{CSR1_vs_CSR2}, using (\ref{CSR_1}) and/or (\ref{CSR_2})
lead to slightly different CSR results, being the former larger
for the $\omega_{max}$-values considered in the experiment. Hence,
the discrepancy between theory and experiment in Fig.~\ref{theory_data} reduces by
$\sim 3\%-5\%$ when Eq.~(\ref{CSR_1}) is also considered within the RMF approach. Further,
for the lowest values of momentum transfer, discrete inelastic excitations of the nuclei may be present in the
data, while they are not considered in the purely QE nucleon knockout estimations of the models.   
\hbox{}From this analysis, we conclude that our theoretical model describes quite consistently 
Bates data, with a minor underestimation (within the experimental error bars), indicating that no 
additional quenching of the relativistic models is needed, other
than the $\sim 10-15\%$ strength that is shifted outside the experimentally available region 
for the $L$ channel. This is consistent with the good agreement found between the RMF model
and the experimental longitudinal scaling function.

The previous argument is also reinforced by results shown in the bottom panels 
of Fig.~\ref{theory_data}. Here we present the ratio of integrated response functions to
the longitudinal strength calculated from the RFG model.
We compare data from Bates
experiment (blue squares) with those given by Meziani et al.~\cite{Saclay1} (green triangles)
and the theoretical results evaluated within the RMF approach (red circles). As in the previous
discussion, the left-bottom panel refers to RMF results evaluated by extending the integral
(\ref{CSR_2}) up to the whole region allowed by kinematics, and dividing by the RFG results. 
This explains why the RMF approach leads to very similar values ($\sim 0.9$) for all $q$, as the comparison
of RMF to RFG results is rather constant with $q$ if integration includes the whole tail region.
On the contrary, in the
right-bottom panel, theoretical RMF-CSR results have been evaluated by using the specific
$\omega_{max}$-limits considered at Bates~\cite{Bates} for each momentum transfer
(red circles). We also show the RMF results obtained by using the momentum transfers $q$
and energy losses $\omega_{max}$ given by Meziani et al.,~\cite{Saclay1} corresponding to
Saclay experiment (black diamonds). Apart from the slightly different $q$ values used
in Bates and Saclay, the difference between 
the RMF results corresponding to both
experimental setups comes from the energy transfer cutoffs considered 
(see Tables~\ref{table1} and~\ref{table2}).
The effect of the cutoff
is particularly visible for $q\sim 475-500$ MeV/c where the larger $\omega_{max}$-values 
considered in Saclay lead to higher RMF-CSR results, making the theoretical prediction 
to depart even further from data.
Therefore, from the general analysis shown in Fig.~\ref{theory_data}, we
observe that RMF calculations are compatible with Bates data in the whole $q$-region,
apart from some deviation (underpredicting data) for the lowest $q=300,325$ MeV/c. On the contrary, 
data from Saclay experiment 
show an important 
depletion ($\gtrsim 40\%$) with regards to the theoretical RMF predictions, even when these data
should include in principle more contribution from the high energy tail (compared with Bates). 
This depletion is not present in Bates data 
and is neither supported by our theoretical estimates. According to the analysis carried out in this work,
this difference in behaviour of Saclay and Bates data cannot be due to 
strength outside the experimental bounds for the energy transfer.

\section{Conclusions}

The study of the CSR and its extraction from the analysis of the separated
$L$ contribution to QE electron scattering data has been extensively
reviewed by different authors leading to rather controversial results. This controversy
is directly linked to the interpretation of experiments as well as to the theoretical
descriptions and the role played by different ingredients. Whereas in some works it is
concluded that a significative quenching occurs in the observed CSR, others show that
only a very mild reduction (or no reduction at all) is observed from the analysis of data. 
Being aware that new, high precision, data expected from Jlab at high energy transfer would help
in disentangling between different approaches, in this work we try to shed some
light on this problem analyzing also its connection with the general scaling properties observed by
inclusive QE electron scattering. 
Scaling arguments applied to $(e,e')$ data have clearly proved
to high accuracy how well scaling is respected by QE data. Moreover,
a ``universal'' superscaling function has been extracted from the analysis of separated
longitudinal data, showing a representative shape with a long tail that extends to high 
values of the energy transfer. As we have shown, this extended tail, with regards to 
usual nonrelativistic models or plane wave approaches, must be kept in mind if 
making estimates of the contribution to the CSR coming from outside the experimentally 
explored region.

A careful analysis has been performed using different theoretical approaches: RPWIA, RMF, rROP.
Results have shown that the CSR is basically independent on the nuclear system considered. Obviously,
for heavy nuclei Coulomb distortion of the electron wave functions would need to be taken into account
in order to extract reliable longitudinal response from the data, but this will not likely affect the
theoretical estimations made in this work. Concerning how the Coulomb 
sum rule reaches its saturation value, we have observed that RPWIA gets saturation faster than RMF. 
This is in accordance with the general shape shown by the superscaling function in both cases, 
being the tail for large $\omega$-values absent in RPWIA. Furthermore, whereas RPWIA leads to a saturated 
CSR very close to 1 for all $q$-values, even when the integration is limited
to the range experimentally considered, the RMF-CSR integrated in the whole allowed range 
gets about $\sim 0.87$ for $q\geq 0.4-0.5$ GeV/c, and this value keeps 
stabilized for $q$ up to the maximum $q\sim 1$ GeV/c explored in this work.

In order to compare our theoretical predictions with experiment, we have analyzed the role
played by the cutoff $\omega$-value considered as upper integration limit in the expression
of the CSR. 
\hbox{}From our results, we conclude that the Coulomb sum rule from RMF reaches 
$\sim 85-95\%$ of its saturated value if truncation at the experimental 
$\omega$-cutoff is taken. The largest strength lost in CSR occurs for $q=475$ MeV/c,
and is of the order of $\sim 15\%$. 
The comparison with CSR results obtained from data measured at Bates for $^{40}$Ca has shown 
its accordance with the RMF approach. Similar comments apply to the ratio of integrated 
response functions to the $L$ strength evaluated with the RFG. These results show 
that no further quenching than the one predicted in the relativistic mean field impulse approximation 
is needed to describe the longitudinal response measured at BATES, that shows a depletion of the free value
of the order of $\sim 10-20\%$. This is in contrast with data measured at Saclay 
showing a reduction of the $L$ channel of the order of $\sim 30-40\%$ \cite{Saclay1,Saclay2,Morg01}, in spite
of the fact that in these experiments the cuttof in the energy transfer is larger than for Bates experiments. 
The reasons for this difference would hopefully be clarified by the recent experiment 
at JLAB~\cite{CSR-JLAB}.

\section*{Acknowledgements}
This work was partially supported by DGI (MICINN-Spain) contracts
FIS2008-04189, FPA2007-62216, by the UCM and Comunidad de Madrid 
(Grupo de F\'{\i}sica Nuclear, 910059), 
the Spanish Consolider-Ingenio programme CPAN (CSD2007-00042), by the
Junta de Andaluc\'{\i}a, and by the INFN-CICYT collaboration
agreements INFN08-20 \& FPA2008-03770-E/INFN.
This work has benefited from discussions with 
M.B. Barbaro and T.W. Donnelly.


\end{document}